\documentclass[11pt]{article}

\usepackage{amsmath}
\usepackage{graphicx}
\usepackage{amsfonts}
\usepackage{amssymb}
\usepackage{epsfig}
\usepackage{color}
\usepackage{psfrag}
\usepackage{epstopdf}

\usepackage{caption}
\usepackage{subcaption}

\newcommand{\EQ}{\begin{equation}}
\newcommand{\EN}{\end{equation}}
\newcommand{\be}{\begin{equation}}
\newcommand{\ee}{\end{equation}}
\newcommand{\bea}{\begin{eqnarray}}
\newcommand{\eea}{\end{eqnarray}}

\begin{document} \setcounter{page}{0}
\topmargin 0pt
\oddsidemargin 5mm
\renewcommand{\thefootnote}{\arabic{footnote}}
\newpage
\setcounter{page}{0}
\topmargin 0pt
\oddsidemargin 5mm
\renewcommand{\thefootnote}{\arabic{footnote}}
\newpage
\begin{titlepage}
\begin{flushright}
\end{flushright}
\vspace{0.5cm}
\begin{center}
{\large {\bf Interface evolution in the two-dimensional quantum Ising model}}\\
\vspace{1.8cm}
{\large Georgios Kampanis$^{1,2}$, Marianna Sorba$^{3}$ and Gesualdo Delfino$^{1,2}$}\\
\vspace{0.5cm}
{\em $^1$SISSA -- Via Bonomea 265, 34136 Trieste, Italy}\\
{\em $^2$INFN sezione di Trieste, 34100 Trieste, Italy}\\
{\em $^3$CNR-INO, Area Science Park, Basovizza, 34149 Trieste, Italy}\\
\end{center}
\vspace{1.2cm}

\renewcommand{\thefootnote}{\arabic{footnote}}
\setcounter{footnote}{0}

\begin{abstract}
\noindent
We consider the unitary time evolution of an interface in the regime of spontaneously broken symmetry of the two-dimensional quantum Ising model. The interface is induced by an initial condition interpolating between the two degenerate ground states in one of the spatial dimensions. The interpolation is left generic in order to investigate the dependence of the late time dynamics on the initial condition. Exploiting the basis of asymptotic quasiparticle states of the bulk theory, the order parameter is analytically determined at large times in the rough phase. The mechanism allowing the breakdown of this phase as the distance from criticality increases emerges from the theory.
\end{abstract}
\end{titlepage}

\newpage
\tableofcontents

\section{Introduction}
Theoretically investigating the dependence of late time nonequilibrium quantum dynamics on its initialization is a particularly nontrivial problem which cumulates the difficulty of determining the evolution at large times with that of handling classes of initial conditions. It was shown in \cite{DS_unitary} how the formalism of asymptotic quasiparticle states allows remarkably general results for the unitary evolution of homogeneous systems in $d$ spatial dimensions: the initial conditions at $t=0$ quantitatively affect the late time dynamics through the values of some amplitudes, while more structural properties such as the presence of oscillations undamped in time \cite{quench} depend on internal symmetries. 

For the case in which the initial conditions introduce spatial inhomogeneities, a first general study could be performed in \cite{q_int} for the one-dimensional quantum ferromagnets in the regime of spontaneously broken symmetry. Considering initial conditions interpolating in an arbitrary way between two different ground states at plus and minus spatial infinity, it was shown that the late time evolution takes place inside a lightcone whose innermost region minimizes the memory of the details of the initial interpolation, reducing it to an overall amplitude for the otherwise universal space-time behavior.

It is the purpose of this paper to extend this type of analysis to the ferromagnetic regime of the two-dimensional quantum Ising model in a transverse field. Now the arbitrary initial interpolation between one ground state at large negative $x$ and the other ground state at large positive $x$ takes place in a strip of width $W_0$ centered around the $y$ axis in the $xy$ plane, so that $W_0$ is the width of an interfacial region at $t=0$. A key technical difference with respect to the one-dimensional case is that the number of quasiparticles whose propagation determines the time evolution of the interfacial region is extensive in the $y$ direction, and is infinite for the infinite $xy$ plane we consider. We will show how this difficulty can be handled and how exact results for the large time evolution are obtained. The region $|x|\ll t$ emerges as that in which the dependence on the initial condition is eventually minimal inside the interfacial region, and the order parameter goes to zero as $1/t$ at any fixed spatial point in the ``rough" phase of the quantum interface in which our derivations are performed. The mechanism allowing the breakdown of the rough phase as the distance from criticality increases also emerges from our complete quantum analysis. 

The paper is organized as follows. In the next section we exploit the basis of asymptotic quasiparticle states for two-dimensional quantum systems with spontaneously broken $\mathbb{Z}_2$ symmetry in order to perform our derivations for the large time evolution of the quantum interface and to characterize interfacial roughening. The case of the Ising model in a transverse field is discussed in section~\ref{ising} before providing some final remarks in Section~\ref{conclusion}.

\section{Quantum interface in two spatial dimensions}
\subsection{One-point functions}
We consider a quantum ferromagnet occupying a plane with coordinates $\mathbf{x}=(x,y)$ and possessing a critical point associated to the spontaneous breaking of a $\mathbb{Z}_2$ spin reversal symmetry. The system is isolated (energy is conserved) and the interactions in its Hamiltonian $H$ do not vary in space. In the spontaneously broken regime of our interest there are two ferromagnetic phases, labeled $+$ and $-$ and interchanged by the symmetry, in which the order parameter (magnetization) takes the values $\langle\sigma\rangle_+>0$ and $\langle\sigma\rangle_-=-\langle\sigma\rangle_+$, respectively. In the vicinity of the quantum critical point the fundamental excitations are quasiparticles with momentum $\mathbf{p}=(p_{x},p_{y})$ and energy $E_{\mathbf{p}}=\sqrt{\mathbf{p}^2+M^2}$, where the mass $M>0$ measures the deviation from criticality. A generic state of the system can be expanded on the basis of asymptotic states $|\textbf{p}_1,\dots ,\textbf{p}_N\rangle$, which are eigenstates of the energy and momentum operators with eigenvalues $\sum_{i=1}^N E_{\mathbf{p}_i}$ and $\sum_{i=1}^N \mathbf{p}_i$, respectively, and for which we adopt the normalization
 \begin{equation}
    \langle \mathbf{q}_1,...,\mathbf{q}_M|\mathbf{p}_1,...,\mathbf{p}_N\rangle = \delta_{M,N} N! \prod_{i=1}^N (2\pi)^2 E_{\mathbf{p}_i} \delta(\mathbf{p}_i-\mathbf{q}_i)\,.
\end{equation}

We want to study the unitary time evolution when the system is initially ($t=0$) in a state which interpolates between the ground state with negative magnetization at $x=-\infty$ and the ground state with positive magnetization at $x=+\infty$. The initial state is taken to be translation invariant in the $y$ direction, and the interpolation in the $x$ direction is kept generic, with the only requirement of left-right symmetry: the magnetization is an odd function of $x$. Both translation invariance in $y$ and left-right symmetry in $x$ will then be preserved by the time evolution. These specifications correspond to the quantum state
\begin{equation}
    |\psi\rangle=\frac{1}{\sqrt{N!}} \int\prod_{i=1}^N \frac{d\textbf{p}_i}{(2\pi)^2E_{\textbf{p}_i}}f(\textbf{p}_1,\dots ,\textbf{p}_N)\,\delta\left(\sum_{i=1}^Np_{y,i} \right)|\textbf{p}_1,\dots ,\textbf{p}_N\rangle\,,
    \label{psi}
\end{equation}
where $N\to\infty$ is understood, since the infinitely long domain wall initial condition necessarily involves infinitely many quasiparticles. The delta function enforces translation invariance in the $y$ direction, while the amplitude $f(\textbf{p}_1,\dots ,\textbf{p}_N)$ depends on the initial condition, namely on the specific interpolation in the $x$ direction. Accordingly, we will perform our analysis for $f(\textbf{p}_1,\dots ,\textbf{p}_N)$ generic, with the only requirement that it decays rapidly enough at large momenta to ensure the convergence of the integrals, is invariant under $p_{x,i}\to -p_{x,i}$ to respect left-right symmetry in the $x$ direction, and is nonsingular at small momenta. 

The expectation value of a local operator $\Phi(\mathbf{x},t)$ is given by
\begin{equation}
    \langle \Phi(\mathbf{x},t)\rangle= \frac{\langle \psi |\Phi(\mathbf{x},t)|\psi\rangle}{\langle \psi|\psi\rangle}\,,
\end{equation}
with normalization
\begin{equation}
    \langle\psi|\psi\rangle = \frac{L}{2\pi} \int\prod_{i=1}^N \frac{d\textbf{p}_i}{(2\pi)^2E_{\textbf{p}_i}} |f(\textbf{p}_1,\dots ,\textbf{p}_N)|^2\, \delta\left(\sum_{i=1}^Np_{y,i} \right)\equiv \frac{L}{2\pi} {\cal N}_f\,,
    \label{norm}
\end{equation}
where we used $\delta(u)=\frac{1}{2\pi}\int dy\, e^{iyu}$ to regularize $\delta(0)=\frac{L}{2\pi}$, with $L\to\infty$ the size of the system in the $y$ direction. We have 
\begin{align}
    &\langle \Phi(\mathbf{x},t)\rangle= \frac{1}{\langle \psi|\psi\rangle N!} \int \prod_{i=1}^N \frac{d\textbf{p}_i}{(2\pi)^2E_{\textbf{p}_i}} \int \prod_{i=1}^N \frac{d\textbf{q}_i}{(2\pi)^2E_{\textbf{q}_i}} f^*(\textbf{p}_1,\dots ,\textbf{p}_N) f(\textbf{q}_1,\dots ,\textbf{q}_N)\,\nonumber\\
    &\times \delta\left(\sum_{i=1}^Np_{y,i} \right) \delta\left(\sum_{i=1}^Nq_{y,i} \right) e^{i \sum_{i=1}^N \left[(p_{x,i}-q_{x,i})x + (E_{\mathbf{p}_i}-E_{\mathbf{q}_i})t\right]} F^{\Phi}_N(\mathbf{p_1},...,\mathbf{p}_N|\mathbf{q}_1,...,\mathbf{q}_N)\,,
    \label{onepoint_function}
\end{align}
where we defined
\begin{equation}
    F^{\Phi}_N(\mathbf{p}_1,...,\mathbf{p}_N|\mathbf{q}_1,...,\mathbf{q}_N) = \langle \mathbf{p_1},...,\mathbf{p}_N|\Phi(0,0)|\mathbf{q}_1,...,\mathbf{q}_N\rangle\,,
    \label{matrix_element}
\end{equation}
and used
\begin{equation}
    \Phi(\mathbf{x},t)=e^{i\mathcal{P}\cdot \mathbf{x}+iH t} \Phi(0,0) e^{-i\mathcal{P}\cdot \mathbf{x}-iH t}\,,
\end{equation}
with $\mathcal{P}$ the momentum operator. The matrix element (\ref{matrix_element}) decomposes into connected and disconnected parts as
\begin{align}
    &F^{\Phi}_N(\mathbf{p}_1,...,\mathbf{p}_N|\mathbf{q}_1,...,\mathbf{q}_N) = F_N^{\Phi,c}(\mathbf{p}_1,...,\mathbf{p}_N|\mathbf{q}_1,...,\mathbf{q}_N) \nonumber\\
    &+ \sum_{i,j=1}^N (2\pi)^2 E_{\mathbf{p}_i} \delta(\mathbf{p}_i-\mathbf{q}_j) F^{\Phi,c}_{N-1}(\mathbf{p}_1,...,\mathbf{p}_{i-1},\mathbf{p}_{i+1},...,\mathbf{p}_N|\mathbf{q}_1,...,\mathbf{q}_{j-1},\mathbf{q}_{j+1},...,\mathbf{q}_N)\nonumber\\
    &+ \sum_{\substack{i,j,l,k=1 \\ i\neq l, j\neq k}}^N (2\pi)^4 E_{\mathbf{p}_i} E_{\mathbf{p}_l} \delta(\mathbf{p}_i-\mathbf{q}_j) \delta(\mathbf{p}_l-\mathbf{q}_k)\nonumber\\
    &\times F^{\Phi,c}_{N-2}(\mathbf{p}_1,...,\mathbf{p}_{i-1},\mathbf{p}_{i+1},...,\mathbf{p}_{l-1},\mathbf{p}_{l+1},...,\mathbf{p}_N|\mathbf{q}_1,...,\mathbf{q}_{j-1},\mathbf{q}_{j+1},...,\mathbf{q}_{k-1},\mathbf{q}_{k+1},...,\mathbf{q}_N)\nonumber\\
    &+ ...\,,
\end{align}
where the connected part is
\begin{equation}
    F_N^{\Phi,c}(\mathbf{p}_1,...,\mathbf{p}_N|\mathbf{q}_1,...,\mathbf{q}_N) = \langle \mathbf{p_1},...,\mathbf{p}_N|\Phi(0,0)|\mathbf{q}_1,...,\mathbf{q}_N\rangle_c\,,
\end{equation}
while the disconnected parts come from annihilations of quasiparticles on the left with quasiparticles on the right, and the final dots indicate the terms with more than two annihilations\footnote{The connectedness structure of matrix elements plays an essential role also in the translation invariant nonequilibrium case, where it determines the lightcone spreading of two-point correlations \cite{2point_spreading}.}.

\subsection{Large time dynamics}
We are interested in the result of the evolution at large times and observe that in this limit the momentum integrals in (\ref{onepoint_function}) are suppressed by the rapid oscillations of the exponential unless the phase is close to the stationarity condition
\begin{equation}
    \nabla_{\mathbf{p}_i}\left(E_{\mathbf{p}_i}t+p_{x,i}x \right)=(v_{x,i}t+x,v_{y,i}t)=0\,,\qquad \qquad i=1,...,N\,,
    \label{stationarity}
\end{equation}
where we introduced the quasiparticle velocities
\begin{equation}
    \mathbf{v}_i=\nabla_{\mathbf{p}_i}E_{\mathbf{p}_i}=\frac{\mathbf{p}_i}{\sqrt{\mathbf{p}_i^2+M^2}}=\frac{\mathbf{p}_i}{E_{\mathbf{p}_i}}\,.
    \label{v}
\end{equation}
Analogous relations are understood for the quasiparticles with momenta $\mathbf{q}_i$. Since $|\mathbf{v}_i|<1$ in our natural units, the stationarity condition (\ref{stationarity}) is satisfied only when $|x|<t$. This means that the time evolution of the one-point function takes place inside the ``lightcone" $|x|<t$. In particular, (\ref{stationarity}) implies that the relevant contribution to the momentum integrals at large time comes from $p_{x,i}$ small for $x$ fixed and $t\gg x$, and from $p_{y,i}$ small\footnote{More generally, the dominant role of low-energy modes in late time nonequilibrium quantum dynamics has been derived for homogeneous \cite{DS_unitary,quench}, inhomogeneous \cite{oscillD}, and long-range \cite{SDD} interactions.} in any case. We can then express the energies as $E_{\mathbf{p}_i}\simeq M+\frac{\mathbf{p}_i^2}{2M}$ and see that each power of momentum contributes a factor $t^{-1/2}$ to the one-point function (\ref{onepoint_function}). Since each annihilation in the matrix element (\ref{matrix_element}) produces a delta function $\delta(\mathbf{p}_i-\mathbf{q}_j)$, and then a factor $t$, the dominant contributions for $t$ large are obtained maximizing the number of annihilations. Denoting by $G^\Phi_j(\mathbf{x},t)$ the contribution to (\ref{onepoint_function}) for $j$ annihilations, the largest $j$ yields
\begin{equation}
    G^{\Phi}_{N}(\mathbf{x},t)=F^{\Phi}_0(|)=\langle 0_-|\Phi|0_-\rangle\equiv\langle\Phi\rangle_-\,,
    \label{G_N}
\end{equation}
where $F^{\Phi}_0(|)$ is the expectation value on the state without quasiparticles, namely a vacuum state, and in the second equality we have specified that we consider the basis of quasiparticle states corresponding to excitations above the vacuum state with negative magnetization. Then the leading non-constant contribution comes from $N-1$ annihilations, so that
\EQ
\langle \Phi(\mathbf{x},t)\rangle\simeq \langle\Phi\rangle_-+G^{\Phi}_{N-1}(\mathbf{x},t)\,,\hspace{1cm}t\to\infty\,.
\label{leading}
\EN
Here and below, $\simeq$ indicates omission of terms subleading at large times. Since $N-1$ annihilations can be performed in $N!N$ possible ways, we have
\begin{align}
    &G^{\Phi}_{N-1}(\mathbf{x},t)= \frac{N}{\langle \psi|\psi\rangle } \int\prod^{N-1}_{i=1}\frac{d\mathbf{p}_i}{(2\pi)^2 E_{\mathbf{p}_i}} \int\frac{d\mathbf{p}}{(2\pi)^2 E_{\mathbf{p}}} \int\frac{d\mathbf{q}}{(2\pi)^2 E_{\mathbf{q}}} f^*(\mathbf{p}_1,...,\mathbf{p}_{N-1},\mathbf{p})\nonumber\\
    &\times  f(\mathbf{p}_1,...,\mathbf{p}_{N-1},\mathbf{q})\, \delta\left(\sum^{N-1}_{i=1}p_{y,i}+p_y\right) \delta(p_y-q_y) \, e^{i[(p_{x}-q_{x})x + (E_{\mathbf{p}}-E_{\mathbf{q}})t]}\, F^{\Phi,c}_1(\mathbf{p}|\mathbf{q})\,.
    \label{G_N-1}
\end{align}

For $t\to\infty$ the last expression involves the matrix element $F^{\Phi,c}_1(\mathbf{p}|\mathbf{q})\biggr\rvert_{\substack{\mathbf{p},\mathbf{q}\to 0\\ p_{y}=q_{y}}}$, since the stationarity condition implies small momenta and the equality of the $y$ components is enforced by the second delta function in the integrand. For scalar operators, $F_1^{\Phi,c}(\mathbf{p}|\mathbf{q})$ can only depend on the single relativistic invariant that can be constructed out of the two momenta. Since the latter can be written as $(\textbf{p}-\textbf{q})^2-(E_\textbf{p}-E_\textbf{q})^2$, and for $p_y=q_y$, $\mathbf{p},\mathbf{q}\to 0$ it reduces to $p_x-q_x$, $F^{\Phi,c}_1(\mathbf{p}|\mathbf{q})\biggr\rvert_{\substack{\mathbf{p},\mathbf{q}\to 0\\ p_{y}=q_{y}}}$ can be expanded in powers of $p_x-q_x$. For the order parameter operator, namely the magnetization operator that we denote by $\sigma$, it is known from \cite{DSS_interface} that this expansion includes a singular term and reads
\begin{equation}
    F_1^{\sigma,c}(\textbf{p}|\textbf{q})\biggr\rvert_{\substack{\mathbf{p},\mathbf{q}\rightarrow 0\\ p_{y}=q_{y}}}=\sum_{k=-1}^{\infty}c^{(k)}_{\sigma}(p_x-q_x)^k\,,
    \label{expansion_matrix_element}
\end{equation}
with
\EQ
c^{(-1)}_{\sigma}=-\frac{2i ML}{N}\,\langle\sigma\rangle_+\,.
\label{c_s}
\EN
Adopting the notation $G^{\sigma,k}_{N-1}$ for the contribution to $G^{\sigma}_{N-1}$ coming from the $k$-th term of the expansion (\ref{expansion_matrix_element}), we have at large time
\begin{align}
    &G_{N-1}^{\sigma,-1}(\mathbf{x},t)\simeq \frac{N}{\langle \psi|\psi\rangle } \int\prod^{N-1}_{i=1}\frac{d\mathbf{p}_i}{(2\pi)^2 E_{\mathbf{p}_i}} \int \frac{dp_x dp_{y}}{(2\pi)^2 E_{\mathbf{p}}} \int \frac{dq_x}{(2\pi)^2 E_{(q_x,p_y)}}\, f^*(\mathbf{p}_1,...,\mathbf{p}_{N-1},\mathbf{p}) \nonumber\\
    &\times f(\mathbf{p}_1,...,\mathbf{p}_{N-1},(q_x,p_y))\, \delta\left(\sum^{N-1}_{i=1}p_{y,i}+p_y\right)\, e^{i[(p_{x}-q_{x})x + (E_{\mathbf{p}}-E_{(q_x,p_y)})t]}\, \frac{c^{(-1)}_{\sigma}}{p_x-q_x}\,.
    \label{G_sing}
\end{align}
Defining $\mu^2=M^2+p_y^2$, we can write $E_{\mathbf{p}}=\sqrt{\mu^2+p_x^2}$ and $E_{(q_x,p_y)}=\sqrt{\mu^2+q_x^2}$, and use the parametrization 
\begin{equation}
    (p_x=\mu\sinh\theta_p,\, E_\textbf{p}=\mu\cosh\theta_p)\,,\quad (q_x=\mu\sinh\theta_q,\, E_{(q_x,p_y)}=\mu\cosh \theta_q)\,.
    \label{rapidity}
\end{equation}
This allows us to rewrite (\ref{G_sing}) as
\begin{align}
    &G_{N-1}^{\sigma,-1}(\mathbf{x},t)\simeq \frac{N c^{(-1)}_{\sigma}}{\langle \psi|\psi\rangle (2\pi)^4 2M} \int\prod^{N-1}_{i=1}\frac{d\mathbf{p}_i}{(2\pi)^2 E_{\mathbf{p}_i}} \int dp_{y}\, d\theta_+\,\frac{ d\theta_-}{\theta_- -i\epsilon}\, e^{2i\mu t\, B\left(\frac{x}{t},\theta_+\right) \sinh\frac{\theta_-}{2}}\nonumber\\
    &\times f^*(\mathbf{p}_1,...,\mathbf{p}_{N-1},\mathbf{p}) f(\mathbf{p}_1,...,\mathbf{p}_{N-1},(q_x,p_y))\,\delta\left(\sum^{N-1}_{i=1}p_{y,i}+p_y\right)\,
\end{align}
where $\theta_{\pm}=\theta_p\pm \theta_q$, 
\begin{equation}
    B\left(\frac{x}{t},\theta_+\right)= \frac{x}{t} \cosh\frac{\theta_+}{2}+ \sinh\frac{\theta_+}{2}\,,
\end{equation}
we took into account that small momenta dominate to write $p_{x}-q_{x}\simeq M\theta_-$, and introduced an $i\epsilon$ prescription for the pole.

We now set $\sinh\frac{\theta_-}{2}=s$ and consider the integral over $s$ in which we close the contour in the upper (lower) complex half-plane if $B$ is positive (negative). In particular, we can close the contour along the line with constant imaginary part $\textrm{Im}\,s=c$. When $t\to\infty$, the contribution coming from the integral on this line is suppressed as $e^{-2\mu|cB|t}$ and can be ignored. On the other hand, we can reduce $|c|$ in such a way that the closed integration contour contains only the singularity at $s=i\epsilon/2$ for $c>0$, and no singularity at all for $c<0$. Cauchy's residue integration then implies that for $t$ large 
\begin{align}
    G_{N-1}^{\sigma,-1}(\mathbf{x},t) &\simeq \frac{2\,\langle\sigma\rangle_+}{{\cal N}_f} \int\prod^{N-1}_{i=1}\frac{d\mathbf{p}_i}{(2\pi)^2 E_{\mathbf{p}_i}} \int \frac{dp_{y}}{(2\pi)^2}\int_{\hat{\theta}}^\infty d\theta_p\,\nonumber\\
    &\times |f(\mathbf{p}_1,...,\mathbf{p}_{N-1},\mathbf{p})|^2\,\delta\left(\sum^{N-1}_{i=1}p_{y,i}+p_y\right)\,,
    \label{G_sing}
\end{align}
where we used (\ref{norm}) and (\ref{c_s}), and $\hat{\theta}$ is the value of $\theta_p$ above which $\tanh\theta_p>-x/t$. Since $\hat{\theta}$ equals $+\infty$ when $x/t<-1$ and $-\infty$ when $x/t>1$, we have
\EQ
G_{N-1}^{\sigma,-1}(\mathbf{x},t)\simeq\left\{
\begin{array}{l}
0\,,\hspace{1.2cm}x<-t\,,\\
\\
2\langle\sigma\rangle_+\,,\hspace{.5cm}x>t\,.
\end{array}
\right.
\label{out}
\EN
On the other hand, $\hat{\theta}\simeq -x/t$ when $|x|/t\ll 1$, so that we can break the integration over $\theta_p$ into that on the small interval between $-x/t$ and $0$, in which $f(\mathbf{p}_1,...,\mathbf{p}_{N-1},\mathbf{p})\simeq f(\mathbf{p}_1,...,\mathbf{p}_{N-1},(0,p_y))$, and that for $\theta_p>0$. Since $f$ is even in $p_x$, the integral for $\theta_p>0$ is half of that for all $\theta_p$, and we obtain
\EQ
G_{N-1}^{\sigma,-1}(\mathbf{x},t)\simeq \left(\mathcal{A}_f \frac{x}{t} +1\right)\langle\sigma\rangle_+\,, \qquad |x|\ll t\,,
\label{deep}
\EN 
where
\begin{equation}
    \mathcal{A}_f=\frac{2}{{\cal N}_f}\int \prod_{i=1}^{N-1} \frac{d\mathbf{p}_i}{(2\pi)^2 E_{\mathbf{p}_i}} \int \frac{dp_y}{(2\pi)^2} |f(\mathbf{p}_1,...,\mathbf{p}_{N-1}, (0,p_y))|^2 \delta \left(\sum_{i=1}^{N-1}p_{y,i}+p_y \right).
\end{equation}

Let us now consider the contribution $G^{\sigma,k}_{N-1}$ that $G^{\sigma}_{N-1}$ receives from a term of the expansion (\ref{expansion_matrix_element}) with $k\geq 0$. Up to a multiplicative constant, this is obtained differentiating $k+1$ times with respect to $x$ the expression (\ref{G_sing}). It then follows from (\ref{out}) that $G^{\sigma,k}_{N-1}$ vanishes for $k\geq 0$ when $|x|>t$. For $|x|\ll t$, it follows from (\ref{deep}) that $G^{\sigma,k}_{N-1}$ vanishes for $k\geq 1$ and would be constant in $x$ for $k=0$. However, since $\langle\sigma(\mathbf{x},t)\rangle$ is an odd function of $x$, also $G^{\sigma,0}_{N-1}$ must vanish, and this implies\footnote{The matrix element $F_1^{\sigma,c}({p}|{q})$ is exactly known \cite{BKW,review} for the transverse field Ising model in one spatial dimension, and has $c_\sigma^{(0)}=0$.} $c_\sigma^{(0)}=0$ in (\ref{expansion_matrix_element}). Taking this into account and recalling (\ref{leading}), (\ref{out}), (\ref{deep}) and $\langle\sigma\rangle_-=-\langle\sigma\rangle_+$, the final result at large times is
\begin{equation}
    \langle\sigma(\mathbf{x},t)\rangle\simeq 
    \begin{cases}
    -\langle\sigma\rangle_+\,, &\qquad x<-t\,,\\
    \mathcal{A}_f\,\langle\sigma\rangle_+\, {x}/{t}\,, &\qquad |x|\ll t\,\\
    \langle\sigma\rangle_+\,, &\qquad x>t\,.
    \end{cases}
    \label{profile}
\end{equation}

\begin{figure}[t]
\centering
\includegraphics[width=7cm]{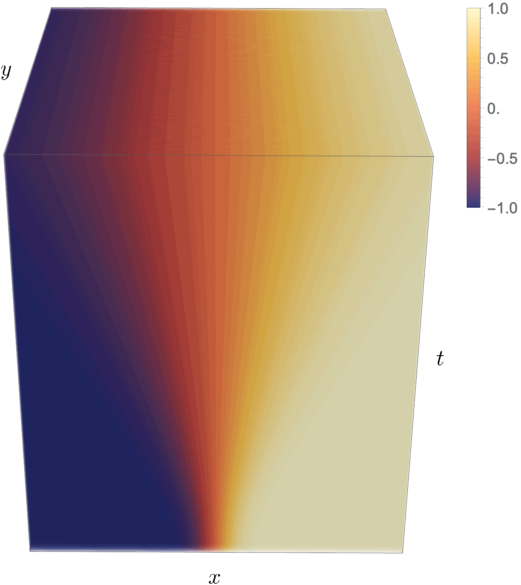}
\caption{Typical appearence of the rescaled magnetization $\langle\sigma(\mathbf{x},t)/\langle\sigma\rangle_+$ in the rough phase, $\langle\sigma\rangle_+$ being the bulk expectation value in the positively magnetized phase.}
\label{lightcone}
\end{figure}

This result has a clear physical interpretation (see also Fig.~\ref{lightcone}). The magnetization profile is constant outside the lightcone, where it takes the two bulk values selected by the initial condition. The location of the lightcone is $|x|\simeq t$ at large times, since the width $W_0$ of the interpolation region in the initial condition becomes anyway negligible as $t\to\infty$. As for the region inside the light cone ($|x|<t$), our derivation made explicit that the generality of the result for $|x|\ll t$ comes from the fact that in this limit the quasiparticles with largest wavelength dominate, and these are maximally insensitive to the details of the spatial modulation of the initial condition. As a consequence, these details only affect the amplitude $\mathcal{A}_f$. On the other hand, when moving from the region $|x|\ll t$ towards the edges of the lightcone, the contribution of the quasiparticles with shorter wavelength becomes more and more important, and the magnetization profile more and more dependent on the function $f$ which encodes the details of the initial condition. In other words, (\ref{profile}) exhibits the universal features of the magnetization profile at large times, namely those features common to the infinite-dimensional space of initial conditions we are considering. 

For any fixed $\bf{x}$, the result
\EQ
\lim_{t\to\infty}\langle\sigma(\mathbf{x},t)\rangle=0
\label{melting}
\EN
follows from (\ref{profile}).

\subsection{Passage probability and roughening}
As we showed, the result (\ref{profile}) is entirely due to the pole term in the expansion (\ref{expansion_matrix_element}). On the other hand, this pole is a kinematical singularity (see e.g. \cite{Smirnov}) independent of the details of the interaction among the quasiparticles, and this is why we were able to derive (\ref{profile}) using only general features of spontaneously broken $\mathbb{Z}_2$ symmetry. This irrelevance of the details of the interaction among the quasiparticles for the large time result (\ref{profile}) can be further understood as follows. The interfacial region, i.e. the region inside the lightcone, is produced by the propagation of $N\to\infty$ quasiparticles distributed along the size $L\to\infty$ of the $y$ direction. In the scaling region close to criticality, which is certainly included in the domain of validity of our derivations in the continuum formalism, the quasiparticle density $N/L$ scales as
\EQ
\frac{N}{L}=\frac{\kappa}{\xi}\,,
\label{density}
\EN
where $\xi=1/M$ is the bulk correlation length and $\kappa$ a dimensionless constant. Since $\xi$ is large in the scaling region, the density is small, meaning that the average distance among the quasiparticles is large and that their mutual interaction is negligible. In the limit towards criticality, in which $\xi\to\infty$, the density vanishes, consistently with the fact that the interface dissolves due to the coalescence of the two ground states. 

\begin{figure}[t]
\centering
\includegraphics[width=7cm]{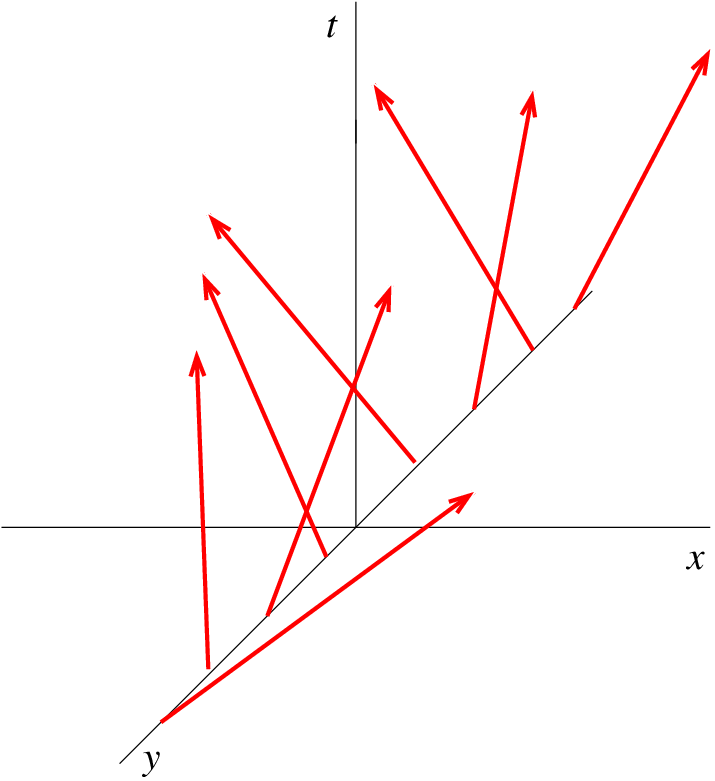}
\caption{Typical configuration of quasiparticle trajectories in the rough phase. The quasiparticles are distributed along the $y$ direction with low density and uncorrelated velocities.}
\label{rough}
\end{figure}

These observations lead to the following probabilistic interpretation of the large time results. A single configuration of the interface is produced by the propagation of quasiparticles distributed along the $y$ axis and traveling with velocities which are uncorrelated due to the small quasiparticle density (Fig.~\ref{rough}). We also know from the stationarity condition that only small values of the $y$ component of the velocities contribute. Eqs.~(\ref{leading}) and (\ref{G_N-1}) show that at large times the order operator $\sigma({\bf x},t)$ couples only to one of the quasiparticles distributed in the $y$ direction. The trajectory $x\simeq v_x t$, with $y$ slowly varying, of this quasiparticle determines the location of the interface around $y$ at time $t$, and we can write
\EQ
\langle\sigma(\mathbf{x},t)\rangle\approx P_L(x,t)\langle\sigma\rangle_+ +(1-P_L(x,t))\langle\sigma\rangle_-=(2P_L(x,t)-1)\langle\sigma\rangle_+\,,
\label{sigma_prob}
\EN 
where $P_L(x,t)$ is the probability that the quasiparticle trajectory passes to the left of $x$ at time $t$, namely that the quasiparticle has velocity $v_x=p_x/E_{\bf{p}}=\tanh\theta_p<x/t$, where we used the parametrization (\ref{rapidity}). The expression of $P_L(x,t)$ is then determined recalling that it follows from (\ref{psi}) that $|f(\mathbf{p}_1,...,\mathbf{p}_{N})|^2$ is, up to normalization, the probability of a state with quasiparticle momenta $\mathbf{p}_1,...,\mathbf{p}_{N}$. The result is proportional to the expression (\ref{G_sing}), namely 
\EQ
P_L(x,t)=\frac{G_{N-1}^{\sigma,-1}(\mathbf{x},t)}{2\langle\sigma\rangle_+}\,.
\label{P_L}
\EN
Recalling that $G_{N-1}^{\sigma,-1}(\mathbf{x},t)-\langle\sigma\rangle_+$ yields (\ref{profile}), we see that (\ref{sigma_prob}) and (\ref{P_L}) consistently provide the probabilistic picture of the interface at large times. Notice that $P_L(x,t)$ actually depends only on $x/t$, since this ratio determines $\hat{\theta}$ in (\ref{G_sing}). In particular, it follows from (\ref{out}) and (\ref{deep}) that at large times
\begin{equation}
    P_L({x},t)=
    \begin{cases}
    0\,, &\qquad x<-t\,,\\
    \frac{1}{2}(1+\mathcal{A}_f\,{x}/{t})\,, &\qquad |x|\ll t\,\\
    1\,, &\qquad x>t\,.
    \end{cases}
\end{equation}
The meaning of this result is clear. The extent of the lightcone in the $x$ direction corresponds to the fluctuations of the interface. The latter cannot pass left of a point located to the left of the lightcone ($x<-t$), and certainly passes left of a point located to the right of the lightcone ($x>t$). For any fixed $x$, the result
\EQ
\lim_{t\to\infty}P_L(x,t)=1/2
\EN
expresses the fact that the width of the interfacial fluctuations in the $x$ direction becomes infinite at infinite time.

Our analytical results hold as long as the bulk correlation length $\xi$ is large enough to keep the quasiparticle density (\ref{density}) small. In this regime the interface is a ``rough" object produced by uncorrelated quasiparticle velocities (Fig.~\ref{rough}). On the other hand, $\xi$ decreases as the distance from bulk criticality increases, and at some point this rough phase should break down because the quasiparticles distributed along the interface get closer and their mutual interaction is no longer negligible. We further discuss this transition in the next section.

\section{Ising model}
\label{ising}
The basic realization of the theoretical setting we have been considering is provided by the two-dimensional quantum Ising ferromagnet in a transverse field, defined by the Hamiltonian
\begin{equation}
    H=-\sum_{\langle i,j\rangle} \sigma_i^x \sigma_j^x -h \sum_i \sigma_i^z\,,
    \label{lattice}
\end{equation}
where $\sigma_i^{x,z}$ are Pauli matrices at site $i$ of a regular planar lattice and the first sum is performed over nearest-neighboring sites; to be definite, we will refer to positive values of the transverse field $h$. At a critical value $h_c$ of the field the model possesses a quantum critical point associated to the spontaneous breaking of spin reversal symmetry in the $x$ direction and belonging to the universality class of the three-dimensional classical Ising model. The spontaneously broken regime corresponds to $h<h_c$ and the order operator denoted so far by $\sigma$ corresponds to the longitudinal magnetization $\sigma^x$. 

Before using the result (\ref{profile}) for the Ising model we have to take into account that it was derived considering the case of a single species of quasiparticles with mass $M_1\equiv M$. On the other hand, for the ferromagnetic phase of the two-dimensional quantum Ising model there is numerical consensus \cite{CH,LSW,CHPZ,DKSTV,Nishiyama,RBLD} on the existence of a second species of stable quasiparticles with mass $M_2\approx 1.8 M_1$, and we need to consider whether this can affect the result for the order parameter profile at large times. While the considerations leading to (\ref{leading}) and (\ref{G_N-1}) do not change, we must include the contributions in which the quasiparticles with momenta ${\bf p}$ and ${\bf q}$ in (\ref{expansion_matrix_element}) are not both of species 1 (i.e. with mass $M_1$). It is clear that these contributions cannot change the result (\ref{profile}) in the region outside the lightcone ($|x|>t$), where no quasiparticle propagates. For $|x|\ll t$, the contribution to $\langle\sigma(\mathbf{x},t)\rangle$ in which both particles are of species 2 will be completely analogous to the one we derived and then proportional to $x/t$. Concerning the off-diagonal contribution, namely that with one quasiparticle of species 1 and the other of species 2, the expansion (\ref{expansion_matrix_element}) will not contain the singular term $k=-1$, since the latter is related (see e.g. \cite{Smirnov}) to the disconnected part associated to the annihilation of the quasiparticle with momentum ${\bf p}$ by that with momentum ${\bf q}$, which can only occur for equal masses. Since we saw that the regular part of (\ref{expansion_matrix_element}) does not contribute to the order parameter profile for $|x|\ll t$, we conclude that the presence of the second quasiparticle species affects (\ref{profile}) only through a redefinition of the amplitude ${\cal A}_f$, so that we have\footnote{While the quasiparticle masses do not appear in (\ref{ising_profile}), it was originally shown in \cite{oscillD} that in spatial dimension $d>1$ a global quench within the ferromagnetic phase of the transverse field Ising model produces persistent oscillations of the spatially constant order parameter with frequencies determined by the quasiparticle masses.} 
\begin{equation}
    \frac{\langle\sigma^x(\mathbf{x},t)\rangle}{\langle\sigma^x\rangle_+}\simeq 
    \begin{cases}
    -1\,, &\qquad x<-t\,,\\
    \tilde{\mathcal{A}}_f\, {x}/{t}\,, &\qquad |x|\ll t\,,\\
    1\,, &\qquad x>t\,.
    \end{cases}
    \label{ising_profile}
\end{equation}
The reason why the region $|x|\ll t$ minimizes the dependence on the initial condition confining it to the amplitude $\tilde{\mathcal{A}}_f$ was explained in the previous section. In particular, for any fixed $\bf{x}$, the central line of (\ref{ising_profile}) implies
\EQ
\lim_{t\to\infty}\langle\sigma^x(\mathbf{x},t)\rangle=0
\label{ising_melting}
\EN
in the regime of validity of (\ref{ising_profile}), i.e. in the rough phase. This phase includes the near critical region, in which the Ising universality class can be identified measuring the critical exponents $\beta\simeq 0.326$ ruling the behavior
\EQ
\langle\sigma^x\rangle_+\sim (h_c-h)^\beta
\EN
of the bulk magnetization entering (\ref{ising_profile}), and $\nu\simeq 0.63$ ruling the behavior
\EQ
M\sim (h_c-h)^\nu
\EN
of the quasiparticle masses. It was shown in \cite{DSS_interface} that the constant $\kappa$ appearing in (\ref{density}) is related to the interfacial free energy of the three-dimensional classical model, and this is why the value $\kappa=0.1084(11)$ \cite{CHP} is numerically known for the Ising universality class. The fact that $\kappa$ is a small number enlarges the range of $h$ in which the density (\ref{density}) is small and the interface is rough.

An interface in the classical three-dimensional Ising model undergoes a roughening transition \cite{BCF,CW} at a temperature $T_r>0$ within the ferromagnetic phase. The phenomenological argument is that the average position of the interface is the $yz$-plane and that a configuration of the interface (ignoring overhangs) is specified by a variable $\phi(y,z)$ measuring the deviation from the average position in the $x$ direction. It is then possible to argue an effective Hamiltonian for $\phi(y,z)$ which allows for a phase transition\footnote{This is no longer the case ($T_r=0$) in one dimension less, due to the absence of transitions in one-dimensional classical systems with short range interactions. Indeed, no roughening transition is exhibited by the interface in the classical two-dimensional Ising model, for which the exact lattice solution is available \cite{Abraham}.}. For the infinite $yz$-plane, the interfacial width $W$ -- i.e. the width of the region spanned by the fluctuations of the interface -- is infinite above $T_r$ (rough phase) and finite below $T_r$ (smooth phase). The divergence as $R\to\infty$ of the interfacial width in the rough phase when the infinite $yz$-plane is replaced by an $\infty\times R$ strip was derived in \cite{DSS_interface}, where the first parameter-free check against Monte Carlo data was also performed. 

In the quantum case, as discussed in the previous section, our analytical results (\ref{ising_profile}) and (\ref{ising_melting}) characterize the rough phase $h_r<h<h_c$, with $h_r$ expected positive to leave room for a smooth phase when the bulk correlation length is no longer large enough to keep the quasiparticle density along the interface small. The interfacial width $W(t)$ corresponds to the width of the lightcone in the $x$ direction at time $t$, and diverges as $t\to\infty$. These results for the rough phase can be generalized to any spatial dimension\footnote{This seems at variance with the conclusion of \cite{Fradkin}, where it was argued that an interface in the zero-temperature quantum Ising ferromagnet is always smooth in $d=3$, while a roughening transition occurs in $d=2$. The criterion for the presence of a rough phase in \cite{Fradkin} is to look for an effective Hamiltonian allowing for a Kosterlitz-Thouless transition. On the other hand, in the classical two-dimensional Ising model the interface is rough despite the absence of a Kosterlitz-Thouless phase.} $d$, and the argument for $h_r>0$ holds as long as there are dimensions longitudinal to the interface, namely in $d>1$. The plots obtained in \cite{ZGEN,EME,HRS} for $d=1$ are consistent with the presence of the rough phase only. Numerical verifications in $d=2$ are substantially more difficult, since sizes and times that can be simulated are still quite limited. Recently, however, a numerical analysis concluding for a transition at $h_r\simeq 0.45\,h_c$ on the square lattice has been presented in \cite{roughening}. On the analytical side, future developments of the small $h$ lattice expansion considered in \cite{Balducci1,Balducci2} might allow to establish a breakdown of (\ref{ising_profile}) and (\ref{ising_melting}) in a smooth phase.

\section{Conclusion}
\label{conclusion}
In this paper we considered the time evolution of an interface in two-dimensional quantum ferromagnets with spontaneously broken $\mathbb{Z}_2$ symmetry, the transverse field Ising model providing the basic realization. The interface is generated by an initial condition at $t=0$ which interpolates from one of the two degenerate ground states at $x=-\infty$ to the other ground state at $x=+\infty$. We generally performed our analysis for the infinite-dimensional space of initial conditions of this type which are left-right symmetric in the $x$ direction (so that $\mathbb{Z}_2$ symmetry is preserved by the initial condition) and translation invariant in the $y$ direction. The analysis was performed within the complete basis of asymptotic quasiparticle states of the the bulk theory and focused on the large time evolution of one-point functions of local operators, in particular the order parameter (longitudinal magnetization for the Ising model). With respect to the one-dimensional case studied in \cite{q_int}, we had to face the additional difficulty that now the interface runs along the direction of the $y$ axis and corresponds to the propagation of infinitely many quasiparticles. In this respect, we showed how the connectedness structure of the matrix elements of the operators on the quasiparticle states play a crucial role and allows to identify the contributions which dominate at large times. The analysis yields the result (\ref{ising_profile}) for the large time longitudinal magnetization in the Ising model. It shows that the interfacial region spreads inside the lightcone $|x|<t$, in which the region $|x|\ll t$ minimizes the dependence on the initial condition, which survives only through the amplitude $\tilde{\mathcal{A}}_f$. This corresponds to the fact that, as we showed, the result for $|x|\ll t$ is determined by the quasiparticles with the longest wavelength, which are then minimally sensitive to the details of the spatial interpolation in the initial condition. Conversely, such details become more and more important when moving towards the edges of the lightcone. On the other hand, for any fixed $x$, the space-time dependence obtained for $|x|\ll t$ holds at sufficiently large times. 

The result (\ref{ising_profile}) holds in the ``rough phase" of the interface, a notion that we showed to naturally emerge from our derivations in the quasiparticle framework. At the same time, the theory also yields the mechanism allowing the rough phase to break down as the distance from criticality increases and the interaction among the quasiparticles distributed along the interface is no longer negligible.

\end{document}